\documentclass[twocolumn]{aastex61}
\usepackage{amsmath,amssymb,amsthm}
\usepackage{wasysym}
\usepackage{bm}
\usepackage{parskip}

\usepackage[normalem]{ulem}

\usepackage{graphicx}                 
\usepackage{xcolor}
\usepackage{tikz}
\usetikzlibrary{fpu,calc,intersections,through,backgrounds}
\usepackage{wrapfig}
\usepackage{MnSymbol}

\received{July 1, 2016}
\revised{September 27, 2016}
\accepted{\today}
\submitjournal{ApJ}

\begin{document}

\title{A new equilibrium state for singly synchronous binary asteroids}

\correspondingauthor{Oleksiy Golubov}
\email{oleksiy.golubov@karazin.ua}

\author[0000-0002-5720-2282]{Oleksiy~Golubov}
\affil{Department of Aerospace Engineering Sciences, University of Colorado at Boulder, 429 UCB, Boulder, CO, 80309, USA}
\affiliation{School of Physics and Technology, V. N. Karazin Kharkiv National University, 4 Svobody Sq., Kharkiv, 61022, Ukraine}
\affiliation{Institute of Astronomy of V. N. Karazin Kharkiv National University, 35 Sumska Str., Kharkiv, 61022, Ukraine}

\author{Vladyslav Unukovych}
\affil{School of Physics and Technology, V. N. Karazin Kharkiv National University, 4 Svobody Sq., Kharkiv, 61022, Ukraine}

\author[0000-0003-0558-3842]{Daniel J. Scheeres}
\affil{Department of Aerospace Engineering Sciences, University of Colorado at Boulder, 429 UCB, Boulder, CO, 80309, USA}

\begin{abstract}

The evolution of rotation states of small asteroids is governed by the YORP effect, nonetheless some asteroids can stop their YORP evolution by attaining a stable equilibrium. The same is true for binary asteroids subjected to the BYORP effect.
Here we discuss a new type of equilibrium that combines these two, which is possible in a singly synchronous binary system. This equilibrium occurs when the normal YORP, the tangential YORP and the binary YORP compensate each other, and tidal torques distribute the angular momentum between the components of the system and dissipate energy.
Such a system if unperturbed would remain singly synchronous in perpetuity with constant spin and orbit rates, as the tidal torques dissipate the incoming energy from impinging sunlight at the same rate.
The probability of the existence of this kind of equilibrium in a binary system is found to be on the order of a few percent.

\end{abstract}

\keywords{minor planets, asteroids: general}

\section{Introduction}
The evolution of asteroids is known to be governed by the non-gravitational effects of sunlight.
In one scenario an asteroid is sped up by YORP to its fission limit, then it settles into a binary, eventually loses its secondary component,
and starts a new YORP cycle.
Still, several equilibria exist along this evolutionary path, which can catch asteroids and thus stop their dynamic evolution.
The diversity of equilibria is produced by the diversity of different manifestations of YORP \citep{vokrouhlicky15}, the most important of which are the normal YORP, or NYORP \citep{rubincam00}, the binary YORP, or BYORP \citep{byorp}, and the tangential YORP, or TYORP \citep{golubov12}

Firstly, a single asteroid can be in equilibrium if its NYORP and TYORP compensate each other \citep{golubov12}.
Secondly, a doubly synchronous binary can be stuck in an equilibrium between NYORP and BYORP \citep{golubov16}.
Thirdly, a semi-equilibrium state exists between BYORP and tides acting on the secondary \citep{jacobson11}, although it does not correspond to a stable state of the primary.

Here we study a more complex and more physically rich equilibrium state, involving all four effects: NYORP, TYORP, BYORP and tides (see the top panel of Figure \ref{fig-equilibrium-explanation}). 
The secondary asteroid resides close to its BYORP-tides equilibrium \citep{jacobson11}.
The primary remains close to its TYORP-NYORP equilibrium \citep{golubov12}, still getting a positive torque from their difference, which compensates the tidal torque created by the secondary. Therefore, we combine the \cite{jacobson11} equilibrium for the secondary with the \cite{golubov12} equilibrium for the primary to produce a global equilibrium.

In Section \ref{sec:theory} we analytically study the simplest model for the effect.
In Section \ref{sec:applications} we consider observational data on asteroid shapes and on the observed properties of binary systems to estimate how realistic the considered equilibrium is.
In Section \ref{sec:discussion} we briefly discuss the implications of this new type of equilibrium.

\section{Theory}
\label{sec:theory}
The simplest model of the phenomenon applies to a primary asteroid with radius $R_1$ and a secondary asteroid with radius $R_2$. Their density $\rho$ is assumed the same, and their mass ratio $q=R_1^3/R_2^3$ is assumed small. The secondary rotates around the primary in a circular orbit of radius $r=R_1a$, which lies in the plane of the system's heliocentric orbit. Rotation axes of both asteroids are perpendicular to this plane. The system is singly synchronous, i.e. the secondary's rotation rate $\omega_2$ coincides with its orbital rotation rate, while the primary's rotation rate $\omega_1$ is different.

The tidal torque created by the primary on the secondary is given by the expression \citep{jacobson11}
\begin{equation}
T_\mathrm{tides} = \frac{2\pi q^2 K_1 \rho R_1^5\omega_d^2}{Qa^6}\mathrm{sgn}(\omega_1-\omega_2).
\label{tides}
\end{equation}
In this equation $Q$ is the quality factor, $K_1$ is the tidal Love number, and $\omega_d=(4\pi G\rho/3)^{1/2}$ is the critical angular velocity of the primary.

The NYORP and BYORP torques are given by \citep{golubov16}
\begin{equation}
T_\mathrm{NYORP} = \frac{C\Phi R_1^3}{c},
\end{equation}
\begin{equation}
T_\mathrm{BYORP} = \frac{B\Phi R_1^3}{c} q^{2/3} a.
\end{equation}
Here $C$ is the NYORP coefficient, $B$ is the BYORP coefficient, $\Phi$ is the solar constant at the asteroid's heliocentric radius, and $c$ is the speed of light.

The approximate expression for TYORP is \citep{golubov17}
\begin{equation}
T_\mathrm{TYORP}=\frac{D\Phi R_1^3}{c}\mathrm{e}^{-\frac{\left(\log{\theta}-\log{\theta_0}\right)^2}{\nu^2}},
\label{TYORP}
\end{equation}
where the constants are $\nu=1.518$, $\log\theta_0=0.580$, and the TYORP coefficient is $D=9\mu n_0$, where $\mu=0.00644$ and $n_0$ characterizes the surface density of boulders. For asteroid 25143 Itokawa $n_0$ is estimated to be $0.028\pm 0.018$ \citep{golubov17,sevecek15}, thus $D=(1.6\pm 1)\cdot 10^{-3}$. The thermal parameter is determined as
\begin{equation}
\label{eq-therm-param}
\theta=\frac{\left(C_s\rho_s\kappa\omega_1\right)^{1/2}}{\left((1-\alpha)\Phi\right)^{3/4}\left(\epsilon \sigma \right)^{1/4}}.
\end{equation}
The physical meaning of the constants used in this equation, as well as their values used for calculations are given in Table \ref{table1}.

The dynamics of the primary and of the secondary are described by the following equations:
\begin{eqnarray}
\label{eq-primary}
I_1\frac{d\omega_1}{dt}=T_\mathrm{TYORP}+T_\mathrm{NYORP}-T_\mathrm{tides}, \\
\label{eq-secondary}
M_2 R_1^2\frac{d(\omega_2 a^2)}{dt}=T_\mathrm{BYORP}+T_\mathrm{tides},
\end{eqnarray}
with $M_2$ being the mass of the secondary and $I_1$ the moment of inertia of the primary. The torques on the right-hand sides should be substituted from Eqs. (\ref{tides})-(\ref{TYORP}), and the orbit's radius can be excluded from the equations via Kepler's third law $a=(\omega_d/\omega_2)^{2/3}$. Thus we arrive at a closed system for two variables, $\omega_1$ and $\omega_2$.

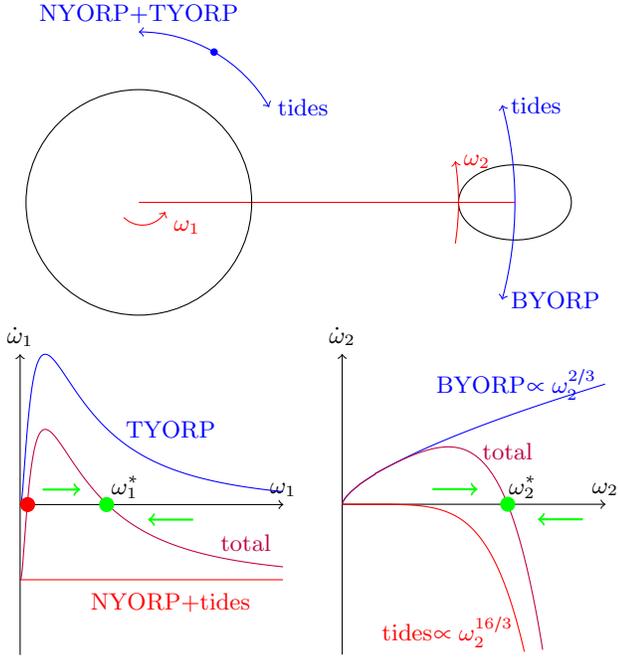
\begin{figure}
\centering
\begin{tikzpicture}[domain=0:3.5]
 \coordinate (A) at (0,5);
 \coordinate (B) at (5,5);
 \coordinate (C) at (5,7);
 \coordinate (D) at (5,3);
 \coordinate (E) at (1,7);
 \coordinate (F) at (4.25,5);
 \coordinate (G) at (-0.2,4.8);
 \draw[-,color=red] (A) -- (B); 
 \draw (A) circle (1.5);
 \draw (B) ellipse (0.75 and 0.5);
 \draw [->,color=red] (F) arc (0:9:3.5) node[right] {$\omega_2$};
 \draw [color=red] (F) arc (0:-9:3.5);
 \draw [->,color=blue] (E) arc (60:90:2) node[above] {NYORP+TYORP}; 
 \draw [->,color=blue] (E) arc (60:30:2) node[right] {tides};
 \fill[blue] (E) circle (0.05);
 \draw [->,color=red] (G) arc (-135:-30:0.35) node[below right] {$\omega_1$};
 \draw [->,color=blue] (B) arc (0:15:5) node[right] {tides};
 \draw [->,color=blue] (B) arc (0:-15:5) node[right] {BYORP};
\end{tikzpicture}
\begin{tikzpicture}[domain=0:3.5]
    \draw[->] (0,0) -- (3.5,0) node[above] {$\omega_1$};
    \draw[->] (0,-2) -- (0,2) node[above] {$\dot{\omega}_1$};
    \draw[->,green,thick] (0.3,0.2) -- (0.8,0.2);
    \draw[->,green,thick] (2.3,-0.2) -- (1.7,-0.2);
    \draw[samples=500,color=red,domain=0:3.5]    plot (\x,-1);
    \node[color=red] at (2,-1.3) {NYORP+tides};
    \draw[samples=500,color=blue,domain=0.00001:3.5]    plot (\x,{2*exp(-(ln(3*\x)/1.5)^2)});
    \node[color=blue] at (2,1.) {TYORP};
    \node[color=purple] at (3,-0.5) {total};
    \draw[samples=500,color=purple,domain=0.00001:3.5]    plot (\x,{2*exp(-(ln(3*\x)/1.5)^2)-1});
    \node[] at (1.4,0.235) {$\omega_1^*$};
    \fill[green] (1.15,0) circle (0.1cm);
    \fill[red] (0.1,0) circle (0.1cm);
\end{tikzpicture}
\begin{tikzpicture}[domain=0:3.5]
    \draw[->] (0,0) -- (3.5,0) node[above] {$\omega_2$};
    \draw[->] (0,-2) -- (0,2) node[above] {$\dot{\omega}_2$};
    \draw[->,green,thick] (1.2,0.2) -- (1.8,0.2);
    \draw[->,green,thick] (3.2,-0.2) -- (2.6,-0.2);
    \draw[samples=500,color=red,domain=0:2.422]    plot (\x,{-14*(\x/3.5)^(16./3.)});
    \node[color=red] at (1.4,-1.7) {tides$\propto\omega_2^{16/3}$};
    \draw[samples=500,color=blue]    plot (\x,{1.6*(\x/3.5)^(2./3.)})    node[left] {BYORP$\propto\omega_2^{2/3}$};
    \draw[samples=500,color=purple,domain=0:2.67]    plot (\x,{1.6*(\x/3.5)^(2./3.)-14*(\x/3.5)^(16./3.)});
    \node[color=purple] at (2.2,0.7) {total};
    \node[] at (2.4,0.235) {$\omega_2^*$};
    \fill[green] (2.2,0) circle (0.1cm);
\end{tikzpicture}
 \caption{Explanation of the equilibrium. \textit{Top:} Sketch of a system capable of YORP-BYORP equilibrium. \textit{Bottom:} Illustration of evolutionary equations for $\omega_1$ and $\omega_2$. Contributions of different torques are shown separately. Stable equilibria are marked by green circles, unstable equilibrium by a red circle. Small green arrows show the direction of evolution in the vicinity of stable equilibria.}
\label{fig-equilibrium-explanation}
\end{figure}

\begin{figure}
 \centering
 \includegraphics[width=0.48\textwidth]{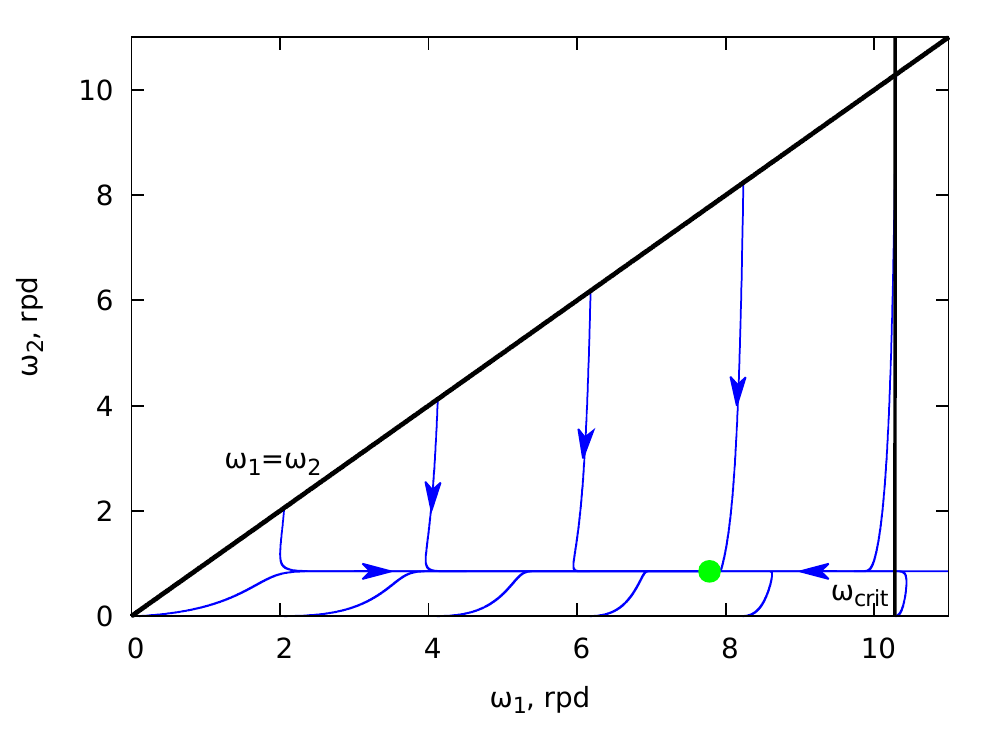}
 \caption{Evolutionary diagram for a binary system with $C=-0.001$, $B=-0.001$ and other parameters as in Table \ref{table1}. The green circle marks the stable equilibrium. Starting from different initial conditions, the system settles down to the the equilibrium.}
 \label{fig-evolution}
\end{figure}

Bottom panel of Figure \ref{fig-equilibrium-explanation} graphically illustrates the possibility of equilibrium in the system described by Eqs. (\ref{eq-primary})-(\ref{eq-secondary}). The right-hand panel shows $\dot{\omega}_2$ as given by Eqn. (\ref{eq-secondary}) in the case $B<0$ and $\omega_1>\omega_2$, which we assume henceforth. We see that in this (and only in this) case a stable equilibrium for $\omega_2$ emerges. This equilibrium $\omega_2$ uniquely determines the tides, thus we treat tides as a constant in the right-hand side of Eqn. (\ref{eq-primary}). If the total torque of NYORP and tides on the primary is slightly negative, then the configuration shown in the left-hand panel of Figure \ref{fig-equilibrium-explanation} can emerge, which has a stable equilibrium. Generally, stability in $\omega_1$ and $\omega_2$ separately does not guarantee stability during their simultaneous evolution. But if $q\ll 1$, then relaxation in $\omega_2$ occurs much faster than in $\omega_1$, and the system rapidly collapses to the equilibrium value of $\omega_2$, and then slowly settles to the equilibrium value of $\omega_1$. An example of such a stable equilibrium is shown in Figure \ref{fig-evolution}, which is constructed by numerically solving Eqs. (\ref{eq-primary})-(\ref{eq-secondary}).

To find the equilibrium in the system, we must equate to zero the right-hand sides of Eqs. (\ref{eq-primary}) and (\ref{eq-secondary}),
\begin{eqnarray}
\label{equil-set}
T_\mathrm{TYORP}+T_\mathrm{NYORP}=T_\mathrm{tides}=-T_\mathrm{BYORP}.
\end{eqnarray}
By substituting Eqs. (\ref{tides})-(\ref{TYORP}) into Eqn. (\ref{equil-set}) and solving the resulting set of equations, we get expressions for the equilibrium dimensionless distance between the asteroids, $a^*$, and the equilibrium thermal parameter, $\theta^*$: 
\begin{align}
\label{solution-a}
a^* =&\Big(-\frac{2\pi c K_1\rho R_1^2\omega_d^2 q^{4/3}}{B\Phi Q}\Big)^{1/7},\\
\theta^* =&\theta_0 \exp\Big(\nu\sqrt{\log\frac{D}{-B q^{2/3}a^*-C}}\Big).
\label{solution-theta}
\end{align}
Then we find the equilibrium value $\omega_1^*$ from Eqn. (\ref{eq-therm-param}) and $\omega_2^*$ from Kepler's law,
\begin{align}
\label{solution-omega1}
\omega_1^*=& \frac{\left((1-\alpha)\Phi\right)^{3/2}\left(\epsilon \sigma \right)^{1/2}\theta^*}{C_s\rho_s\kappa},\\
\omega_2^*=& \omega_d (a^*)^{-3/2}.
\label{solution-omega2}
\end{align}
Thus, given the YORP coefficients $B$, $C$, and $D$, we can define the equilibrium spins and separation of the system. Eqn. (\ref{solution-theta}) gives the real value of $\theta^*$ only if the following conditions are met:
\begin{equation}
\label{YORP-cond}
-D<B q^{2/3}a^*+C<0,
\end{equation}
This equation means that the curve for the total torque in the bottom right panel of Figure \ref{fig-equilibrium-explanation} is neither too high nor too low, but has an intersection with 0. This condition can also be thought of as a generalization of the more special conditions for the BYORP-tides equilibrium $B<0$ \citep{jacobson11} and the TYORP-NYORP equilibrium $-D<C<0$.

Several more conditions should be met for the formal solution of Eqn. (\ref{equil-set}) to be a reasonable equilibrium. Firstly, the primary should rotate faster than the secondary (otherwise no stable equilibrium is possible), but slower than the critical rotation rate (otherwise a rubble-pile primary gets disrupted by the centrifugal forces),
\begin{equation}
\label{omega-cond}
\omega_2<\omega_1<\omega_d.
\end{equation}
Secondly, the radius of the secondary's orbit should be larger than the primary's radius, but smaller that the Hill limit,
\begin{equation}
\label{a-cond}
1<a<a_\mathrm{max},
\end{equation}
(See \cite{golubov16} for the definition of $a_\mathrm{max}$.) Still, these conditions are not very restrictive: $1<a$ necessarily follows from Eqn. (\ref{omega-cond}), while $a_\mathrm{max}$ is never reached in our simulations, as discussed in the following section.

\section{Applications}
\label{sec:applications}

\begin{figure*} 
	\centering
	\includegraphics[width=0.48\textwidth]{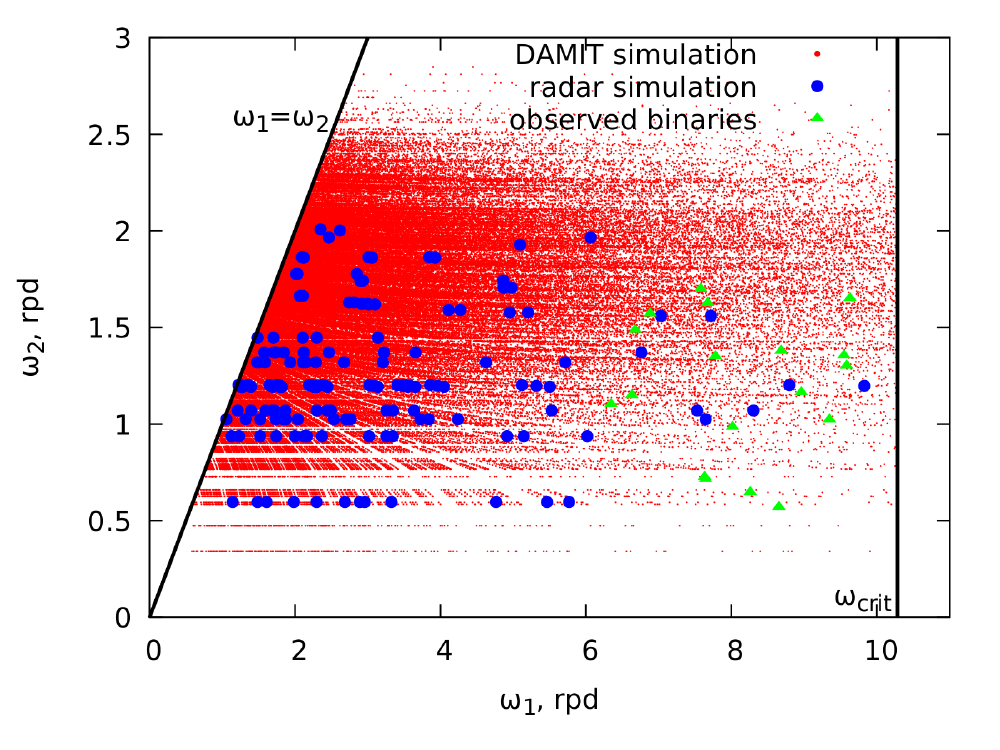}   
	\includegraphics[width=0.48\textwidth]{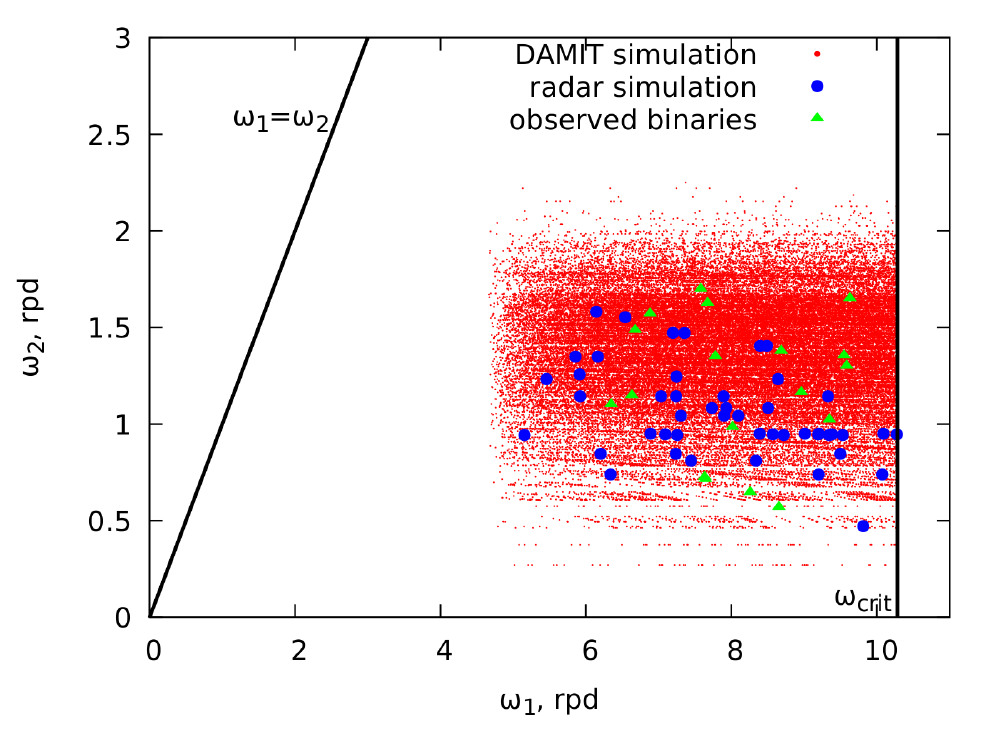}   
	\includegraphics[width=0.48\textwidth]{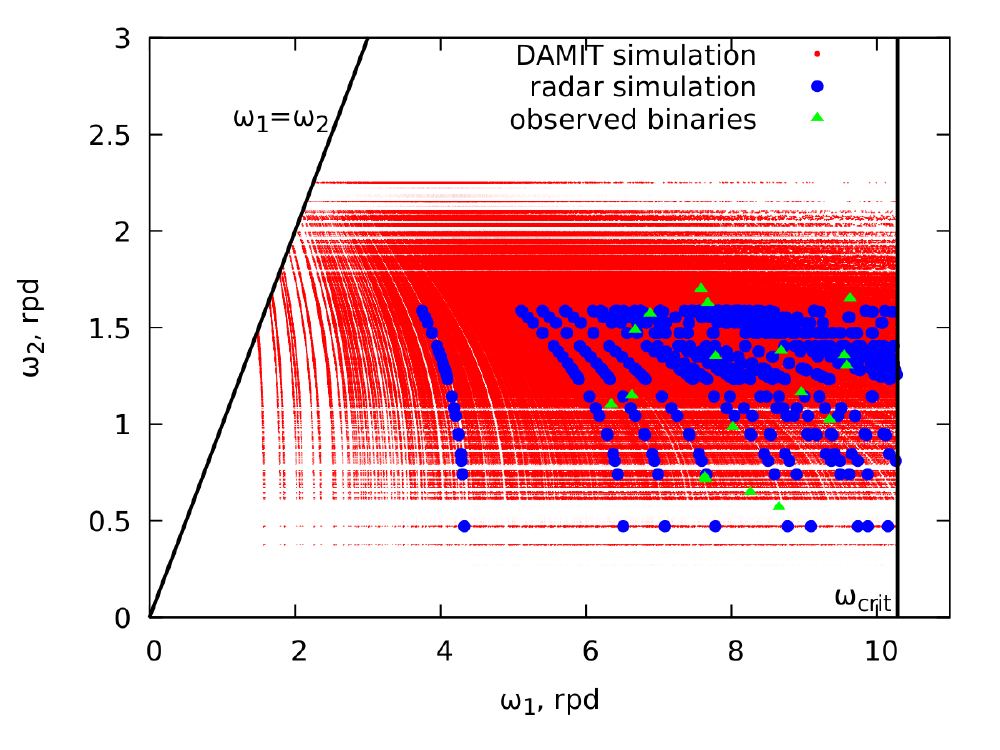}
	\includegraphics[width=0.48\textwidth]{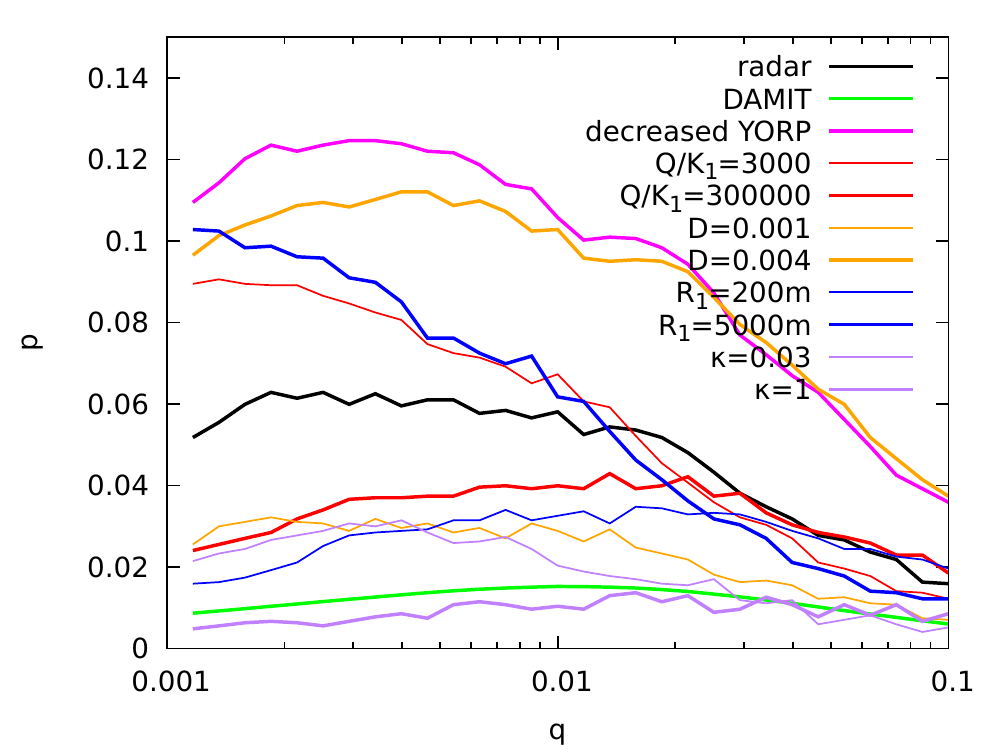}
	\caption{Probability of equilibria. All values of parameters are taken from Table \ref{table1}, if not stated otherwise. \textit{Top left:} Distribution of the equilibrium rotation states over $\omega_1$ and $\omega_2$. Simulations for the DAMIT and radar shapes with the standard values of the relevant parameters (Table \ref{table1}) are overplotted with the observed distribution in binaries.
	\textit{Top right:} The same, but with $Q/K_1=10000$ and $\kappa=0.03$ $\mathrm{{W}\,m^{-2}K^{-1}}$. 
	\textit{Bottom left:} The same, but with $Q/K_1=1000000$ and all values of $C$ and $B$ divided by 100.
	\textit{Bottom right:} Probability of equilibrium as a function of the mass ratio $q$. Different curves correspond to different values of relevant parameters. The first two lines are calculated for the radar and DAMIT shapes with the standard values of the relevant parameters from Table \ref{table1}. The third line (``decreased YORP'') is computed for the YORP coefficients $C$ and $B$ twice smaller than in the radar database. For the remaining six lines the radar database is used, and all parameters are standard (Table \ref{table1}), except of one parameter that is altered.}
	\label{fig-probability}
\end{figure*}

\begin{table}
\caption{Constant standard values}
\begin{tabular}{ c | c | l }
\hline
Notation & Value & Meaning\\
\hline
$q$ & $0.01$ & mass ratio\\
$R_1$ & $1000$ $\mathrm{m}$ & primary's radius\\
$R_2$ & $R_1 q^{-2/3}$ & secondary's radius\\
$\omega_1$ & & primary's rotation rate\\
$\omega_2$ & & secondary's rotation rate\\
$\omega_d$ & $(4\pi G\rho/3)^{1/2}$ & critical rotation rate\\
$a$ & $(\omega_d/\omega_2)^{3/2}$ & distance between the com-\\
& & ponents in terms of $R_1$\\
\multicolumn{3}{c}{\textit{Dynamical properties}}\\
$Q/K_1$ & $30000$ & ratio of the quality factor\\
&  & to the tidal Love number\\
D & $0.002$ & TYORP constant\\
$\rho$ & $2$ $\mathrm{\frac{g}{cm^3}}$ & density of the asteroids\\
\multicolumn{3}{c}{\textit{Thermal properties}}\\
$\Phi$ & 1361 $\mathrm{\frac{W}{m^2}}$ & solar constant at the\\
& & asteroid's position\\
$\rho_s$ & $2.5$ $\mathrm{\frac{g}{cm^3}}$ & density of stones\\
$C_s$ & $680$ $\mathrm{\frac{J}{kg\,K}}$ & heat capacity of stones\\
$\kappa$ & $0.26$ $\mathrm{\frac{W}{m^{2}K^{1}}}$& heat conductivity of stones\\
$\alpha$ & 0.1 & albedo\\
$\epsilon$ & 0.9 & thermal emissivity\\
$\sigma$ & $5.67\cdot 10^{-8}$ $\mathrm{\frac{W}{m^2 K^4}}$ & Stefan-Boltzmann constant\\  
\hline
\end{tabular}
\label{table1}
\end{table}

To estimate the importance of this kind of equilibrium for real asteroids, we study two different sets of asteroid shape models: photometric shape models from the DAMIT database constructed by lightcurve inversion technique \citep{damit} and radar shape models \citep{radar}. As there can be a statistical difference between shapes of single and binary asteroids, the use of such shapes can somewhat bias our results. In addition we show spin and orbit periods from observed binary asteroids \citep{pravec16} to assess whether the theoretical values we find are consistent with the population. 

We take all possible pairs of asteroid shapes from our databases, assuming one of them to be the primary and the other to be the secondary, compute their YORP coefficients $C$ and $B$ \citep{NYORP16,golubov16}, and check each pair for the equilibrium using Eqs. (\ref{solution-a})-(\ref{a-cond}). For all the physical parameters we take the standard values as given in Table \ref{table1}, thus we rescale the shape models of the primary and the secondary to the new size assuming the fixed values for the radius of the primary $R_1$ and the mass ratio $q$. The resulting equilibrium angular velocities $\omega_1$ and $\omega_2$ in all the pairs are plotted in the top left panel of Figure \ref{fig-probability}. The area where the equilibria lie is limited by the two black lines, which correspond to Eqn. (\ref{omega-cond}). On the other hand, Eqn. (\ref{a-cond}) is not restrictive, as the condition $a=a_\mathrm{max}$ is indistinguishable from $\omega_2=0$. With green triangles, we overplot $\omega_1$ and $\omega_2$ for the confirmed or probable singly synchronous binaries from \cite{pravec16}. We see a substantial overlap between our simulation and the observational data, although the observed binaries tend to avoid slow  $\omega_1$. 

The agreement between the simulation and the observations can be much improved by alteration of values of certain physical parameters from Table \ref{table1}. In the top right panel of Figure \ref{fig-probability}, the heat conductivity $\kappa$ is decreased. Such a small heat conductivity could correspond to TYORP produced by the regolith \citep{golubov17}. In the bottom left panel of Figure \ref{fig-probability}, all assumed values of $B$ and $C$ are divided by 100. This assumption could better reproduce binary asteroids, which can be more symmetric than single asteroids from DAMIT and radar shape databases. The values of $Q/K_1$ are also altered in both cases (decreased in the top right panel and increased in the bottom left panel of Figure \ref{fig-probability}). If the discussed kind of equilibrium is widely spread between binary asteroids, then adjustment of parameters for the best agreement between the simulations and the observations could be used to constrain the uncertain physical properties of asteroids, such as $\kappa$ and $Q/K_1$.

Then we use the same sets of asteroid shape models to compute the percentage of asteroid pairs, in which our simulations predict existence of an equilibrium as a function of the assumed mass ratio between the shapes, and plot the probability of equilibrium in the bottom right panel of Figure \ref{fig-probability}. The black line shows the radar shape models database with the parameters as given in Table \ref{table1}, the green line -- the same for the DAMIT database. The other lines correspond to the radar database with only one parameter varied with respect to Table \ref{table1} and the other parameters left unaltered.  
From the bottom right panel of Figure \ref{fig-probability} we see that the equilibrium probability is sensitive to parameter variations, but always remains on the order of a few percent.

\section {Discussion}
\label{sec:discussion}

We have found a new kind of equilibrium of a singly synchronous binary asteroid system. In this equilibrium, the secondary asteroid resides in an equilibrium between BYORP and tides (similar to \cite{jacobson11}), while the primary has an equilibrium between TYORP, NYORP and tides (reminiscent to \cite{golubov12}, but with tides added). 

In this system, radiation torques input angular momentum to the primary and take away the equivalent angular momentum from the secondary. Tides serve as a link for transporting angular momentum from the primary to the secondary. Tidal friction permanently consumes energy, but the energy of the system is perpetually supplanted by the mechanical work performed by radiation torques.

The probability of the existence of this equilibrium is found to be on the order of a few percent, and is thus comparable with the previously estimated probabilities of TYORP-NYORP equilibria of single asteroids \citep{golubov12} and NYORP-BYORP equilibria of doubly synchronous asteroids \citep{golubov16}. This mechanism can extensively exclude asteroids from their YORP-cycles and lock them in stable equilibria in the form of singly-synchronous binaries. Such effects will strongly bias the observed statistics of binary asteroid spin rates.

\end{document}